\documentclass[adp,fleqn]{w-art}
\usepackage{times}
\usepackage{w-thm}

\usepackage[]{graphicx}
\chardef\bslash=`\\ 

\begin{document}

\DOIsuffix{theDOIsuffix}

\Volume{12}
\Issue{1}
\Copyrightissue{01}
\Month{01}
\Year{2003}

\pagespan{1}{}


\title[QCD Sum Rules and Models for GPDs]{QCD Sum Rules and Models for Generalized \\
Parton Distributions}

\author[Radyushkin]{Anatoly Radyushkin\footnote{E-mail: 
{\sf radyush@jlab.org}. Also at Laboratory of Theoretical Physics, Dubna, 
141980, Russian Federation}\inst{1,2}}
\address[\inst{1}]{Physics Department, Old Dominion University,  Norfolk, VA 23529, USA}
\address[\inst{2}]{Theory Group, Jefferson Lab, Newport News, VA 23606, USA}
 \dedicatory{Dedicated to Klaus Goeke on occasion of his 60th birthday}

\begin{abstract}
  
I use  QCD sum rule ideas 
 to   construct  models for 
generalized parton distributions.  
To this end, the perturbative parts of  QCD sum rules  
for the pion and nucleon electromagnetic form factors
are  interpreted in terms of GPDs and two models are discussed.
One of them takes the double Borel transform at adjusted value of the Borel 
parameter as a model for nonforward parton densities,
and another is based on the local duality relation. 
Possible ways  of improving these Ans{\"a}tze are briefly discussed.

\end{abstract}
\maketitle

\section{Introduction}
\label{sect1}

 The concept of Generalized Parton 
 Distributions \cite{Muller:1998fv,Ji:1996ek,Radyushkin:1996nd,Radyushkin:1997ki} 
  is a  modern  tool to provide a more detailed description
  of the hadron structure.  The need for GPDs is 
  dictated by 
   the present-day 
 situation  
in hadron physics, namely:     
$i)$ The fundamental particles 
from which the hadrons  are built are known:   quarks and gluons. 
 $ii)$ Quark-gluon interactions are described by   
 QCD whose   Lagrangian is also  known. 
$iii)$ The knowledge of these first principles
is not sufficient at the moment, and we still  need 
hints from experiment to  understand how QCD works, and we need to translate  
 information 
obtained on the hadron level 
into the language  of 
quark and gluonic fields. 

One can consider  projections of  combinations of quark and gluonic fields
onto hadronic states 
$|P \rangle$ 
$ \langle  \, 0 \, | \, \bar q_\alpha (z_1) \, q_\beta (z_2) \, 
| \, P \, \rangle $, etc., 
      and  interpret them  as hadronic wave functions.
         In principle, solving  the 
  bound-state  equation $H  | P \, \rangle =  E | P \,\rangle $
   one should get   
   complete
  information about the hadronic structure.
  In practice,  the equation
  involving infinite number of Fock components has never been solved.
  Moreover, the  wave functions are not directly accessible 
  experimentally.
   The way out is to  use  phenomenological functions. 
  Well known examples are form factors, 
   usual parton densities, and distribution amplitudes. 
 The new functions, 
   Generalized Parton Distributions 
   \cite{Muller:1998fv,Ji:1996ek,Radyushkin:1996nd,Radyushkin:1997ki} 
   (for   recent reviews,  
   see \cite{Goeke:2001tz,Diehl:2003ny}),   
  are  hybrids of these ``old'' functions which, in their turn,  are the 
  limiting cases of  the ``new''  ones.

\section{Generalized parton distributions}
\label{sect2}

Generalized parton distributions parametrize nonforward matrix 
elements of lightcone operators. 
For, example, the twist-2 part of the  vector operator built of quark fields 
${\cal O}^\mu (z)=\bar \psi (-z/2) \gamma^\mu \psi (z/2)$  in the simplest case 
of a (pseudo)scalar hadron, e.g., pion  can be parametrized in two ways.
The first one is in terms of the off-forward 
parton distribution \cite{Muller:1998fv,Ji:1996ek,Radyushkin:1996nd}
\begin{equation}
\langle P-\frac{r}{2} \, | \, {\cal O}^\mu (z) \,|\, P+\frac{r}{2} \rangle = 
2P^\mu \int_{-1}^1 dx \, e^{-ix(Pz)} 
H(x,\xi,t) \  , 
\end{equation}
(where $\xi = (rz)/2 (Pz)$  is the skewness of the matrix element and $t=r^2$)  
or in terms of two double distributions (DDs) 
\cite{Muller:1998fv,Radyushkin:1996nd,Polyakov:1999gs}
\begin{equation}
\langle P-\frac{r}{2} \, | \, {\cal O}^\mu (z) \,|\, P+\frac{r}{2} \rangle =
\int_{-1}^1 d\beta \int_{-1+|\beta|}^{1-|\beta|} d\alpha \, e^{-i\beta(Pz)-i\alpha (rz)/2} 
\bigl \{2P^\mu F(\beta,\alpha,t) +r^\mu G(\beta,\alpha,t) \bigr \} \ . 
\end{equation}
The variables $x,\xi$ of OFPDs ($\beta,\alpha$ of DDs) 
can be  interpreted   
as momentum fractions: initial  and  returning quarks 
carry the momenta $(x+\xi)P^+$ and $(x-\xi)P^+$, ($\beta P^+ +(1+\alpha)r^+/2$
and $\beta P^+ -(1-\alpha)r^+/2$), respectively. 
The functions $H(x,\xi,t)$, $F(\beta,\alpha,t)$, $G(\beta,\alpha,t)$ are related by
\begin{equation}
H(x,\xi,t) = \int_{-1}^1 d\beta \int_{-1+|\beta|}^{1-|\beta|} d\alpha
\delta (x-\beta -\xi \alpha)\bigl \{F(\beta,\alpha,t) +\xi G(\beta,\alpha,t) \bigr \} \  .
\end{equation}
The resolution of the apparent discrepancy of describing
the same object in terms of one or two functions is based on the observation 
that 
the choice of two DDs $F, G$ is not unambiguous
\cite{Polyakov:1999gs,Teryaev:2001qm}: one can perform transformations
which do not change  the combination $\partial F/\partial \beta + \partial G/\partial \alpha$ 
\cite{Teryaev:2001qm}.
In particular, there exists a DD representation 
in terms of a single function \cite{Belitsky:2000vk}
\begin{eqnarray}
\langle P-\frac{r}{2} \, | \, {\cal O}^\mu (z) \,|\, P+\frac{r}{2} \rangle &=&
\int_{-1}^1 d\beta \int_{-1+|\beta|}^{1-|\beta|} d\alpha \, e^{-i\beta(Pz)-i\alpha (rz)/2} 
\bigl \{2\beta P^\mu +\alpha r^\mu \bigr \} h(\beta,\alpha,t)\nonumber \\
&=&
2i\, \frac{\partial}{\partial z_\mu} 
\int_{-1}^1 d\beta \int_{-1+|\beta|}^{1-|\beta|} d\alpha \, e^{-i\beta(Pz)-i\alpha (rz)/2} 
 h(\beta,\alpha,t) \ .
\end{eqnarray}
The generalized parton distribution functions provide a 
very detailed description of the   hadronic structure.
They include information contained in simpler 
functions, like  usual parton densities $f(x)$ and form factors $F(t)$, 
reducing to them in particular  limits. The forward limit gives $H(x, \xi=0, t=0) =f(x)$,
while the  local  one produces the reduction formula 
\begin{equation}
\int_{-1}^1 H(x,\xi,t) \, dx = F(t) \ .
\end{equation}
Equivalent relations between DDs, parton densities and form factors also can be  written.

Intermediate in   complexity    are  nonforward parton densities \cite{Radyushkin:1998rt}
 ${\cal F} (x,t) = H(x, \xi=0,t )$, or GPDs at zero skewness.
 They reduce to parton densities for zero $t$, and give form factors after 
 integration over $x$. 
 The functions ${\cal F} (x,t)$ can be also obtained 
 from the $F$-DDs by integration over $\alpha$:
 \begin{equation}
 {\cal F} (x,t) = \int_{-1+|x|}^{1-|x|} 
 F(x,\alpha,t) \, d\alpha \  .
\label{DDtoND}
 \end{equation}
 Note, that the $\alpha$-integral of $G$-DDs is zero \cite{Bakulev:2000eb,Tiburzi:2002tq} 
 because they are odd functions of $\alpha$. 
 Interplay between $x$ and $t$ dependence of 
 ${\cal F} (x,t)$ is an interesting and nontrivial problem.  
  In particular, it is closely associated with 
 the question \cite{Drell:1969km}
of interrelation between large-$t$ behavior of hadron
 form factors and $x\to 1$ shape of parton densities.
  
 GPDs accumulate information 
 about long-distance interactions, hence, they are   nonperturbative functions. 
 Possible ways to get  theoretical estimates for them
 include lattice QCD \cite{Negele:2004iu}
 and QCD-inspired models
 \cite{Ji:1997gm,Petrov:1998kf,Diehl:1998kh,Praszalowicz:2003pr}.
Building  self-consistent  models of GPDs is, however,  a rather 
 difficult problem, because one needs to satisfy many 
 constraints which should be obeyed by GPDs. They include   
  spectral properties,
  polynomiality condition, positivity, relations
 to parton densities and form factors \cite{Muller:1998fv,Ji:1996ek,Radyushkin:1996nd},
 soft pion theorems \cite{Polyakov:1998ze,Kivel:2002ia}. 
 Most of  these conditions are satisfied, of course, in  
 perturbation theory, and there were attempts 
  \cite{Mukherjee:2002gb,Pobylitsa:2002vw}
 to use 
 perturbative expressions for modeling GPDs. 
 Still, there remains a question 
 about relation of perturbative results
 to functions describing nonperturbative dynamics.
 In this respect, QCD sum rules \cite{Shifman:1978bx}
look  
 as an attractive possibility, being  an  approach 
 which is closely related to  Feynman diagrams.  
  Its basic concept,
  quark-hadron duality, provides a tool 
 for translating perturbative results into statements about 
 nonperturbative functions. In the past,  
 QCD sum rules were used  to get 
 information about form factors 
 \cite{Ioffe:1982qb,Nesterenko:1982gc,Nesterenko:1983ef,Bakulev:1991ps,Bakulev:2000uh}
and parton densities (see \cite{Ioffe:2002gt} and references  therein,
and \cite{Belitsky:1996vh}). 
A natural idea 
 is to apply the 
QCD sum rule techniques to model  GPDs.
This idea in the pion case was already elaborated in  some detail
 by the Bochum/Dubna  group  \cite{Bakulev:2000eb}.
My goal in the present paper  is  to 
give interpretation of the approach in terms 
of double distributions,  to discuss the 
 nucleon case, and  to combine 
the idea  with recent developments.

\section{QCD Sum Rule for Pion Form Factor}
\label{sect3}

Basic objects of QCD sum rule analysis \cite{Shifman:1978bx} are  correlators of local
currents  with  quantum numbers of the hadrons one intends to study.
The usual choice for the pion is the axial current
$
j_{\alpha} = \bar d \gamma_5 \gamma_{\alpha} u
$. Its projection on the 
single-pion state $\langle 0|j_{\alpha}(0)|p \rangle = i f_{\pi} p_{\alpha}$
is specified by the pion decay constant $f_{\pi}$.
To study  the pion form factor,  one should consider  
correlator of  three currents \cite{Ioffe:1982qb,Nesterenko:1982gc}
\begin{equation}
T^{\mu}_{\alpha\beta}(p_1,p_2) =
i \int e^{-ip_1z_1+ip_2z_2} 
\langle 0\, |\, T\{j_{\beta}(z_2) J^{\mu}(0) j^+_{\alpha}(z_1)\} 
|\,0\, \rangle d^4z_1 d^4z_2 ,
\end{equation}
where 
$J^{\mu} = e_u\bar u \gamma^{\mu} u + e_d \bar d \gamma^{\mu} d$
is the electromagnetic current.  
The pion-to-pion transition  term corresponds to 
$$
\langle 0|j_{\beta}(z_2)|p_2\rangle \langle p_2 |J^{\mu}(0)|p_1 \rangle 
\langle p_1| j^+_{\alpha}(z_1)| 0 \rangle \ , 
$$
where  the pion form factor  contribution 
$$
\langle p_2 |J^{\mu}(0)|p_1 \rangle = 2P^{\mu} F_{\pi}(Q^2)
$$
appears in the middle matrix element.  As usual, $Q^2=-q^2$ or 
$Q^2=-t$, if the GPD notation  $t$ is used.  
The relevant invariant amplutude can be extracted  by 
taking the projection \cite{Nesterenko:1982gc}
$$T^{\mu}(p_1,p_2) \equiv n^{\alpha}n^{\beta} 
T^{\mu}_{\alpha\beta}(p_1,p_2)/(nP)^2
  \ , 
$$
 with $n^{\alpha}$ chosen to be a lightlike  vector  
with equal projections on $p_1$ and $p_2$,  $(np_1) =(np_2)\equiv (nP)$. 
Since $(nq)=0$ and $n^2=0$,  the projection  kills 
the structures containing $q_\alpha, q_\beta $ and $g_{\alpha\beta}$. 
Still, the projection may contain $n^\mu$ terms, which 
cannot be directly related to the form factor contribution.
 In what follows, we will  always omit them without explicit notice.

 The starting point  of  the QCD sum approach \cite{Shifman:1978bx} is the dispersion 
 relation for  
invariant amplitudes that appear in the correlator, in fact,  the double 
dispersion relation in case of a three-point function \cite{Ioffe:1982qb,Nesterenko:1982gc},
\begin{equation}
T(p_1^2,p_2^2,Q^2)=\frac1{\pi^2}\int_0^\infty ds_1\int_0^\infty ds_2
 \ \frac{\rho(s_1,s_2,Q^2)}{(s_1-p_1^2)(s_2-p_2^2)} \ .
\label{eq:doubledr}
\end{equation}
One should  find  the  expression for  the perturbative version of the correlator
and  nonperturbative 
corrections which modify the spectrum in the $p_i^2$ channels,  
converting  the free-quark spectral density into
a  function  containing  physical hadrons.  
From a practical point of view,
 it is more convenient to consider the double Borel transform
 \cite{Ioffe:1982qb,Nesterenko:1982gc}
\begin{equation}
\Phi(\tau_1,\tau_2,Q^2) =\frac1{\pi^2}\int_0^\infty ds_1\int_0^\infty ds_2
 \ \rho(s_1,s_2,Q^2) \, e^{-s_1 \tau_1 -s_2 \tau_2}
\label{eq:doublebor}
\end{equation}
in which power weights are substituted by the exponential 
ones. Formally, the action 
of the Borelization operator 
 is given by  $${\cal B}(p^2 \to \tau) \{1/(s-p^2)\} = e^{-s\tau}.$$
  QCD sum rule for the pion form factor then has the structure
 \begin{eqnarray} 
   f_\pi^2 F_\pi (Q^2)e^{-m_\pi^2(\tau_1+\tau_2)} &+& {\rm higher \ states} 
 \nonumber \\  &=& \frac{1}{\pi^2} \int_0^{\infty}ds_1 \int_0^{\infty}ds_2\, 
    \rho^{\rm pert}(s_1,s_2,Q^2) \, e^{-s_1\tau_1 -s_2 \tau_2}
    \nonumber \\  &+&
   \tau^2 A(\tau_i/\tau) \langle \alpha_s G^2 \rangle +\tau^3 B(\tau_i/\tau,Q^2) 
   \,  \alpha_s \langle \bar qq \rangle^2
   \nonumber \\ &+& {\rm higher \ condensates} \  ,
    \label{fffull}
   \end{eqnarray} 
where $\tau \equiv \tau_1 +\tau_2$. 
We kept the pion mass on the left hand side,  but 
it will be neglected  from now on.  
  
\section{ Structure of perturbative term  }
\label{sect4}

In the  case of  $T^{\mu}(p_1,p_2)$ amplitude, the lowest order
perturbative term is given by a triangle diagram.
It is very convenient to write it in the   $\alpha$-representation
(some details can be found in Refs. \cite{Radyushkin:1997ki,Efremov:1978cu})  
\begin{eqnarray} 
T^{\mu\, {\rm pert}} (p_1, p_2) = \frac{3}{2\pi^2} \int_0^{\infty} 
\frac{1}{\lambda^2}
\, d \alpha_1d \alpha_2 d \alpha_3
\frac{\alpha_3(\alpha_1+\alpha_2)}{\lambda^2}  
\left \{ 2 P^\mu \frac{\alpha_3}{\lambda} + q^\mu \frac{\alpha_1- \alpha_2}{\lambda}
\right \} 
\nonumber \\
\exp \left [ \frac{-Q^2\alpha_1 \alpha_2 +p_1^2 \alpha_1\alpha_3 + p_2^2 \alpha_2\alpha_3}
{\alpha_1 + \alpha_2 + \alpha_3}  \right ]  \  , 
\end{eqnarray}
where $\lambda = \alpha_1 + \alpha_2 + \alpha_3$. 
Using the formula \cite{Nesterenko:1982gc,Nesterenko:1983ef}
$$ {\cal B}(p_i^2 \to \tau_i)\{ e^{A_i p_i^2} \}=  \delta (\tau_i - A_i) $$  
one can obtain  the  double Borel transform
$$\Phi^\mu (\tau_1,\tau_2,Q^2) \equiv {\cal B}(p_1^2 \to \tau_1){\cal B}(p_2^2 \to \tau_2)
T^\mu (p_1^2, p_2^2, Q^2)$$ of the perturbative   amplitude  
\begin{eqnarray} 
\Phi^{\mu\, {\rm pert}}(\tau_1,\tau_2,Q^2) 
= \frac{3}{2\pi^2}  \int_0^{\infty} 
\frac{1}{\lambda^2}
\, d \alpha_1d \alpha_2 d \alpha_3 \, 
\frac{\alpha_3(\alpha_1+\alpha_2)}{\lambda^2} 
\left \{ 2 P^\mu \frac{\alpha_3}{\lambda} + q^\mu \frac{\alpha_1- \alpha_2}{\lambda}
\right \} 
\nonumber \\
 \delta \left (\tau_1 -   \frac{\alpha_1\alpha_3}{\lambda} \right ) 
\delta \left (\tau_2 -   \frac{\alpha_2\alpha_3}{\lambda}\right )   \ 
\exp \left [-Q^2 \frac{\alpha_1 \alpha_2}{\lambda} \right ] . 
\end{eqnarray}
 It is instructive to rewrite this expression using  new variables 
$$
x= \frac{\alpha_3 }{ \lambda}   \ ; \  
\frac{\alpha_1}{ \lambda} = \rho (1-x) \equiv \rho \bar x \  ; 
\  \frac{\alpha_2}{ \lambda} = (1-\rho)  \bar x  \equiv \bar \rho \bar x \  .
$$
This gives the following integral representation 
\begin{eqnarray} 
\Phi^{\mu\, {\rm pert}}(\tau_1,\tau_2,Q^2) 
= \frac{3}{2\pi^2}  \int_0^{\infty} 
d \lambda 
\,\int_0^1 \bar x dx \int_0^1 d \rho \ x(1-x)  
\left \{ 2 x  P^\mu +  
\bar x (\rho - \bar \rho)  q^\mu 
\right \} 
\nonumber \\
 \delta \left (\tau_1 - \rho \lambda   x \bar x \right ) 
\delta \left (\tau_2 -   \bar \rho \lambda  x \bar x \right)   \ 
\exp \left [-\lambda \rho \bar \rho \bar x^2 Q^2  \right ] \  . 
\end{eqnarray}
Two delta functions can be used to perform 
integration over $\lambda$ and $\rho$. The result is    
\begin{equation} 
\Phi^{\mu\, {\rm pert}} (\tau_1,\tau_2,Q^2) =  \frac{3 }{ 2\pi^2 (\tau_1 + \tau_2)} 
\int_0^1 \bar x \, dx \,  \left \{2 x  P^\mu + \bar x q^\mu 
\frac{\tau_1 - \tau_2}{\tau_1 + \tau_2} \right \} 
\exp \left [ - Q^2\frac{(1-x) \tau_1 \tau_2}{x(\tau_1+\tau_2)} \right ]  \  . 
 \label{eq:Phipert2}
\end{equation}

\section{GPD interpretation of perturbative term}

 In case of  GPDs, we should substitute the local current by the
bilocal operator ${\cal O}_\mu (z)$, in which 
quark fields are separated by a lightlike distance
$z$, and  then use parametrization in terms of OFPDs or DDs.
In fact, nothing prevents us from  taking  $z$ 
equal to the projection vector $n$. 
Then calculation of the triangle diagram contribution 
completely parallels that for the purely scalar case  
discussed in Refs. \cite{Radyushkin:1997ki,Mukherjee:2002gb}.
 The conversion to DD variables is especially transparent.
 One should just use the fact  that the spinor factors ${\not k}_1, {\not k}_2 $
  corresponding to the numerators 
 of quark propagators 
 adjacent to the composite vertex can 
 be written in the $\alpha$-representation 
  with vectors $k_1, k_2$ given by 
 $${ k_1} = { p_1} \frac{\alpha_3}{\lambda} +{ r}  
 \frac{\alpha_2}{\lambda} = 
 P \frac{\alpha_3}{\lambda} + \left ( 1+  \frac{\alpha_2- \alpha_1}{\lambda} \right ) 
 \frac{ r}{2} 
 \  , \ 
 { k_2} = { p_2} \frac{\alpha_3}{\lambda} - { r}  
 \frac{\alpha_1}{\lambda} = 
 P \frac{\alpha_3}{\lambda} -  \left ( 1-  \frac{\alpha_2- \alpha_1}{\lambda} \right ) 
 \frac{{r}}{2}  \  ,
 $$
 while the corresponding quark momenta in DD variables are
 $\beta P +(1+\alpha)r/2$ and $\beta P - (1 - \alpha)r/2$,
 respectively.   
 Thus, $\alpha_3/\lambda$ should be interpreted 
 as $\beta$, and $(\alpha_2- \alpha_1)/\lambda$
 as $\alpha$. This mnemonics helps to understand
 the DD  representation of the triangle diagram 
\begin{eqnarray} 
{\cal T}^{\mu\, {\rm pert}} (p_1, p_2;\alpha,\beta) = \frac{3}{2\pi^2} \int_0^{\infty} 
\frac{1}{\lambda^2}
\, d \alpha_1d \alpha_2 d \alpha_3 \, 
\delta \left (\beta - \frac{\alpha_3}{\lambda} \right )
\delta \left (\alpha - \frac{\alpha_2-\alpha_1}{\lambda} \right )
\nonumber \\ \times  
\frac{\alpha_3(\alpha_1+\alpha_2)}{\lambda^2}  
\left \{ 2 P^\mu \frac{\alpha_3}{\lambda} + r^\mu \frac{\alpha_2- \alpha_1}{\lambda}
\right \} 
\exp \left [ \frac{-Q^2\alpha_1 \alpha_2 +p_1^2 \alpha_1\alpha_3 + p_2^2 \alpha_2\alpha_3}
{\alpha_1 + \alpha_2 + \alpha_3}  \right ]  \  . 
\end{eqnarray}
It differs from the representation for $T^\mu (p_1,p_2)$ by two 
delta functions relating the DD variables $\alpha,\beta$ to the
$\alpha$-parameters.
Note also, that defining  GPDs we   treat the momentum transfer
$r=p_1-p_2$ as ``going  upwards'' in the $t$-channel,
i.e., we take  $r=-q$.
 Using the delta functions to eliminate  two integrations, we obtain
  \begin{eqnarray} 
{\cal T}^{\mu\, {\rm pert}} (p_1, p_2;\alpha,\beta) = \frac{3\, \theta (\beta) }{4\pi^2} 
\beta (1-\beta) \,\left \{ 2 \beta P^\mu  + \alpha r^\mu 
\right \} 
\int_0^{\infty} 
d \lambda 
\, e^{-\frac14 \lambda   Q^2[(1-\beta)^2 -\alpha^2]} 
\nonumber \\  
\times \exp \left \{ 
\frac12 \lambda \beta [(1 - \beta - \alpha   )p_1^2
+ (1 - \beta + \alpha  )p_2^2
 ]  \right \}  \  . 
\end{eqnarray}
The restriction $\beta >0$ reflects the obvious fact  that the triangle diagram
involves only   valence quarks.  
Now it is straightforward to calculate the double Borel transform
   \begin{eqnarray} 
\Phi^{\mu\, {\rm pert}} (\tau_1, \tau_2;\alpha,\beta) &=& 
\frac{3\, \theta (\beta) }{2\pi^2(\tau_1+\tau_2)} 
 \, (1-\beta) \left \{ 2 \beta P^\mu  + \alpha r^\mu 
\right \} 
\nonumber \\  &\times & 
\delta \left [\alpha - \frac{\tau_2-\tau_1}{\tau_1+\tau_2}(1-\beta) \right ]
\, e^{-  Q^2(\tau_1 + \tau_2) \frac{(1-\beta)^2 -\alpha^2}{4\beta(1-\beta)}} 
\  . 
\end{eqnarray} 
  Note, that the spectral property $|\alpha|\leq 1-\beta$ is manifest
  in this expression.

\section{``Borel'' model for nonforward densities}

Integrating DD over $\alpha$, we should get  nonforward 
parton density (see Eq.(\ref{DDtoND})). Performing the integral, we obtain
\begin{eqnarray} 
\int_{-1+\beta}^{1-\beta}
 \Phi^{\mu\, {\rm pert}} (\tau_1, \tau_2;\alpha,\beta) \, d\alpha &=& 
 \frac{3\, \theta (\beta) }{2\pi^2(\tau_1+\tau_2)} 
 \, (1-\beta) \left \{ 2 \beta P^\mu  +   (1-\beta)\, r^\mu\frac{\tau_2-\tau_1}{\tau_1+\tau_2} 
\right \} 
\nonumber \\  &\times & 
\, \exp \left [ -  \frac{(1-\beta)Q^2\tau_1  \tau_2 }{\beta(\tau_1+\tau_2)} \right ] 
\  . 
\end{eqnarray} 
Comparing this result with Eq.(\ref{eq:Phipert2}), we see that the integration 
variable $x$ in that equation has the meaning of the momentum 
fraction, and the integrand can be treated as a 
 nonforward parton density  ${\cal F} (x,t=-Q^2)$.  

Writing QCD sum rule for the pion  form factor,
we should treat symmetrically both pions, the initial and the final one,
i.e.  take 
 the   Borel parameters $\tau_1,\tau_2$  equal to each other, as 
 it was done 
in Refs.\cite{Ioffe:1982qb,Nesterenko:1982gc}. Such a choice corresponds to DD 
$\Phi^\mu$ proportional to  $ \delta (\alpha)$.
In other words, each  quark takes exactly a half  of 
the $t$-channel momentum $r$: there is no spread in the distribution
of $r$  among constituents.  
Furthermore, the $G$ part of DD   vanishes for $\tau_1=\tau_2$.
As a result, the perturbative term for the off-forward distribution 
in this case 
\begin{eqnarray}
H^{\rm pert}_a(x,\xi,t)|_{\tau_1=\tau_2\equiv \tau}
=6N_a  x(1-x) \, 
\exp \left [   \frac{(1-x)}{2x}t\tau \right ] 
\label{gauss}
\end{eqnarray} 
coincides with that for 
the nonforward parton density, or   in the forward $t=0$ limit, 
with the perturbative term for the usual parton density $f(x)$. 
The perturbative term suggests  $6x(1-x)$ for the shape of  the 
normalized  parton density. This does not look realistic,
even for a valence distribution. However,
as shown in Ref. \cite{Belitsky:1996vh}, nonperturbative corrections
shift the maximim of the distribution to smaller $x$, 
and then  DGLAP evolution from a low normalization 
point $\mu^2 \sim 0.25 \,$GeV$^2$ 
(to  which  QCD sum rules refer) produces acceptable 
valence distributions for the pion. 
In general, it is a rather popular  idea 
  that there are no skewness  effects
at a low normalization
point, and one can start 
with the forward approximation,  generating 
nontrivial $\xi$ dependence through evolution \cite{Frankfurt:1997ha,Martin:1997wy}.
 
 Our result (\ref{gauss}) gives an example of a nontrivial
 interplay between $x$ and $t$ dependence of a nonforward parton density.
 It has the same  form as the result
 of calculation of  the overlap contribution 
 of two lightcone wave functions 
 $\Psi(x,  k_{\perp})$ with the Gaussian $\sim \exp[-a^2k_{\perp}^2/x(1-x)]$
 dependence on transverse momentum. 
   The parallel between the Borel transform and Gaussian wave functions
   is well known (see, e.g., Refs. \cite{Nesterenko:1983ef,Radyushkin:1995pj}).  
   It can be  explained 
   by the exponential weight $e^{-s\tau}$ and the lightcone form 
   $s= (k_{\perp}^2+m_q)^2/x(1-x)$ for the invariant mass of the
   $\bar qq$ system. This analogy with 
   the wave function description suggests to take the Borel transform
   $\Phi^{\mu\, {\rm pert}} (\tau, \tau;\alpha,\beta)$ 
  at a particular (adjusted) value 
   of  the Borel parameter   as a model for the pion DD,  
   treating $\tau$ as 
   the width scale   of a Gaussian wave function.
    From the QCD sum rule   point of view, 
   the model  
   \begin{eqnarray} 
   f_\pi^2 F_\pi^B (Q^2) =\frac{1}{\pi^2} \int_0^{\infty}ds_1 \int_0^{\infty}ds_2\, 
    \rho^{\rm pert}(s_1,s_2,Q^2) \, e^{-(s_1 +s_2) \tau}
    \label{FBorel}
   \end{eqnarray} 
    corresponds to taking such a value of $\tau$ for which    
   the condensate corrections and  the subtraction of contributions 
   due to higher states perfectly cancel.
   For each fixed   $Q^2$,  such a value of $\tau$ exists,
   but in principle it may  depend  on $Q^2$.
   The absence of such a dependence can be expected only
   if the description  
   of the pion vertex by a Gaussian wave function 
   is a very good approximation of  reality.
   However, inside the QCD sum rule approach,
   the  wave function backing of this model  faces  difficulties.
   In particular,  
    one can try to check      the normalization
   condition $F_\pi (0)=1$ by 
   using   the Ward identity relation 
 $$ \rho^{\rm pert}(s_1,s_2,Q^2=0) = \pi \delta (s_1-s_2) 
 \rho^{\rm pert}(s_1) \ . $$ 
  between  three-point and two-point function densities. 
  The resulting expression 
 \begin{eqnarray} 
   f_\pi^2 F_\pi^B (0)=\frac{1}{\pi} \int_0^{\infty} 
    \rho^{\rm pert}(s) \, e^{-2s\tau} \, ds
   \end{eqnarray} 
   matches the expression for  $f_\pi^2$
   derived from   the  two-point  sum rule
 \begin{eqnarray} 
   (f_\pi^B)^2 =\frac{1}{\pi} \int_0^{\infty} 
    \rho^{\rm pert}(s) \, e^{-s\tau} \, ds
   \end{eqnarray} 
 only if  
   one decreases the $\tau$  parameter 
   of the three-point function by factor 2 compared to that used 
   in the two-point function relation.

  \section{Local quark-hadron duality model}

   Another approach  to obtaining  predictions for 
   hadronic characteristics from   their analogues 
   calculated  for 
   free-quark systems  is suggested by the local quark-hadron duality 
   hypothesis.
    For the pion form factor, it gives \cite{Nesterenko:1982gc}
    \begin{eqnarray} 
   f_\pi^2 F_\pi^{LD} (Q^2) =\frac{1}{\pi^2} \int_0^{s_0} ds_1 \int_0^{s_0}ds_2\, 
    \rho^{\rm pert}(s_1,s_2,Q^2) \ ,
    \label{FLD}
   \end{eqnarray}  
 where $s_0$ is the duality interval.  
 This relation corresponds to the $\tau=0$ limit 
 of the full QCD sum rule (\ref{fffull}) 
 in which the  higher 
 states are modeled by the perturbative spectral density starting 
 at $s_0$ in both $s_1$ and $s_2$ directions.   
 When  $\tau \to 0$,  the condensate corrections 
 vanish,  and the exponential weight 
 $e^{-(s_1+s_2)\tau}$ converts into 1. 
 The local duality approach  has no problems 
 with the $Q^2=0$ limit. It  gives  for $f_\pi^2 F_\pi^{LD} (0)$ 
 the same expression 
 as the local duality relation
  \begin{eqnarray} 
   (f_\pi^{LD})^2 =\frac{1}{\pi} \int_0^{s_0} 
    \rho^{\rm pert}(s) \, ds
   \end{eqnarray} 
based on two-point sum rule gives for $f_\pi^2$.  Numerically, the result 
$\rho^{\rm pert}(s) = 1/4\pi$ 
\cite{Shifman:1978bx}
fixes $s_0$ at $4\pi^2 f_\pi^2 \approx 0.7$\,GeV$^2$ \cite{Nesterenko:1982gc}.
This value is also the result \cite{Shifman:1978bx} of fitting full QCD sum rule, with 
condensates included.
 
 The duality interval $s_0$ has the meaning of the effective 
 threshold for the onset of higher states.
 Since the location of higher states  is fixed, $s_0$ has 
 good chances to be $Q^2$-independent.
   This hope is supported by the fact that the 
   local duality prediction  \cite{Nesterenko:1982gc,Radyushkin:1990te}
 for the pion form factor
     \begin{eqnarray} 
\left (1+\frac{\alpha_s}{\pi} \right )
F_\pi (Q^2) = \left ( 1- \frac{1+6s_0/Q^2}{(1+4s_0/Q^2)^{3/2}} \right ) + 
\frac{\alpha_s/\pi}{1+Q^2/2s_0}
\label{LDpion}
\end{eqnarray} 
(with $\alpha_s/\pi =0.1$ and $s_0=4\pi^2 f_\pi^2$)
 is in perfect agreement with the latest Jefferson Lab 
measurements \cite{Volmer:2000ek}.  
 The ${\cal O}(\alpha_s)$ term in Eq.(\ref{LDpion})  
   is the simplest interpolation \cite{Radyushkin:1990te} between the 
   $Q^2=0$ value fixed by the Ward identity and the large-$Q^2$ 
   asymptotic behavior  $F_\pi (Q^2) \to 8\pi \alpha_s f_\pi^2/Q^2$ 
    \cite{Efremov:1978rn,Lepage:1980fj} 
   due  to the one-gluon exchange.

Knowing the double Borel transform one can obtain  the spectral
density  
by the inverse transformation. It is convenient to write the result 
 in a form similar to the lightcone representation \cite{disser}  
\begin{eqnarray}
 \rho^{\rm pert}(s_1,s_2,Q^2) =
 \frac{3}{2 \pi}
 \int_0^1 x \bar x \, dx \int d^2 \kappa_{\perp} 
\delta (s_1 -\kappa_{\perp}^2) \, \delta (s_2 -(\kappa_{\perp}+\tilde q_{\perp})^2) \ , 
\label{density} 
\end{eqnarray} 
where $\tilde q_{\perp}$ is a two-dimensional vector with
$\tilde q_{\perp}^2= \bar x Q^2/x$.
From this representation, it is evident that 
the ``Borel'' model (\ref{FBorel}) corresponds to Gaussian 
wave functions, while the local duality (\ref{FLD})
corresponds to step-like effective wave functions
 $\sim  \theta (\kappa_{\perp}^2 \leq s_0)$
 (note, that $\kappa_{\perp}$  differs from the usual
 $k_{\perp}$ by $\sqrt{x\bar x}$ rescaling).
Thus,  to get the form factor  through the local duality 
 formula, one needs to calculate the area of  overlap of
 two circles having equal radii. This gives the representation \cite{disser} 
  \begin{eqnarray}
 F_\pi^{LD} (Q^2) = \frac{12}{\pi} \int^1_{1/(1+4s_0/Q^2)} x \bar x \, dx 
 \left \{ \arccos \sqrt{\frac{\bar xQ^2}{4xs_0}}-
 \sqrt{\frac{\bar xQ^2}{4xs_0}\left (1-\frac{\bar xQ^2}{4xs_0} \right ) } \right \}  \ .
\end{eqnarray} 
Its  integrand  can be treated as 
the local duality model for the nonforward parton density ${\cal F} (x, t=-Q^2)$. 
 Simple structure of Eq. (\ref{density}) allows to write 
 form factor also in the impact parameter representation
  \begin{eqnarray}
   F_\pi^{LD} (Q^2) = \frac{6}{\pi} \int \frac{d^2 b_{\perp}}{ b_{\perp}^2}
   e^{i (q_{\perp}b_{\perp})}
   \int^1_0 x \bar x  \, \left [ J_1 \left (\sqrt{\frac{xs_0}{1-x}}b_{\perp}
   \right ) \right ] ^2\, dx
  \ ,
\end{eqnarray} 
where $J_1(z)$ is the Bessel function.
Again, the integrand gives a model for the nonforward density in the
impact parameter space.

\section{Nucleon form factors}

QCD sum rule analysis of the nucleon form factors  
is based on the study of  the 3-point correlator 
\begin{eqnarray}
T^{\mu}_{\alpha \beta} (p_1,p_2)=\int
\langle 0\,|\,T\{ \eta_\beta (z_2) J^\mu (0) \bar \eta_\alpha (z_1)\} \,|\,0\rangle
e^{-i(p_1z_1)+i(p_2z_2)} d^4z_1 d^4z_2
\label{1}
\end{eqnarray}
of  the electromagnetic current $J^\mu$ 
and two Ioffe currents $\eta, \bar \eta$ \cite{Ioffe:1981kw} 
with   the nucleon
quantum numbers,   
$$
\eta = \varepsilon^{abc} \left(u^a{\cal C}^{-1} \gamma_\rho
u^b\right)\gamma_\rho\gamma_5d^c \ .
$$
Here, ${\cal C}$ is the charge conjugation matrix, 
$\{a,b,c\}$ refer to quark colors,  the absolutely antisymmetric
tensor $\varepsilon^{abc}$ ensures that the  currents are
color singlets,  and $\alpha,\beta$ are Dirac indices.
The amplitude $T^{\mu}_{\alpha \beta} $ 
is the sum of various structures:
$P^\mu {\not \!P}\equiv V^\mu(P), \ 
q^\mu {\not \!P}, \ 
i\epsilon^{\mu \lambda \rho \sigma} P_\lambda q_\rho \gamma_5 \gamma_\sigma
\equiv A^\mu(P,q),\
Q^2 \gamma^\mu$, etc. To compare contributions of different structures,
one should specify the reference frame.
A natural choice is the infinite momentum frame (IMF), where
$P^\mu $ is in the plus  direction and $P^+ \to \infty$, while
$q^\mu \equiv q^\mu_{\perp}$ is fixed.  
Note, that neglecting 
$q^\mu$ compared to $P^\mu$ is exactly what we
did in the pion case by taking projection on the
lightlike vector $n$ orthogonal to $q$.
The leading IMF
structure is clearly $V^\mu(P)$: it does not contain the
``small''  parameter $q$. 
For $p_1^2=p_2^2$, this structure 
satisfies the transversality condition $q_\mu V^\mu(P)=0$.
Another structure possessing this property is $A^\mu(P,q)$.
These two structures have the  most direct connection 
with the ${\not\!\!P}$ component of the two-point
function defined through the correlator
$\langle 0\,|\,T\{ \eta_\beta (z)  \bar \eta_\alpha (0)\} \,|\,0\rangle$. 
In the decomposition of the 
proton-to-proton transition part of 
the correlator  $T^{\mu}_{\alpha \beta} (p_1,p_2)$,
the structure $V^\mu(P)$  is accompanied by the nucleon 
$F_1 (Q^2)$ form factor, while the $A^\mu(P,q)$ structure
is  accompanied by the form factor $G_M(Q^2)$. 
The double Borel transform of the invariant amplitude
$T_V(p_1^2,p_2^2,Q^2)$ related to  
$V^\mu(P)$ structure has the form \cite{Nesterenko:1983gi}
\begin{eqnarray} 
\Phi_1(\tau_1,\tau_2,Q^2) 
= \frac{1}{(2\pi)^4(\tau_1+\tau_2)^3}  
\int_0^1 dx \, \left [3e_u \bar x^2 -(2e_u-e_d)\bar x^3 \right ]  \,
 \exp \left [ - Q^2\frac{\bar x \tau_1 \tau_2}{x(\tau_1+\tau_2)} \right ] \,  .
\end{eqnarray}
Taking $\tau_1=\tau_2\equiv \tau$, one obtains the Borel model
for the $F_1(Q^2)$ form factor. Similarly, the double Borel
transform of the $T_A(p_1^2,p_2^2,Q^2)$ amplitude \cite{Nesterenko:1983gi}
\begin{eqnarray} 
\Phi_M(\tau_1,\tau_2,Q^2) 
= \frac{1}{(2\pi)^4(\tau_1+\tau_2)^3}  
\int_0^1 dx \ 3e_u \, \bar x^2 \, 
 \exp \left [ - Q^2\frac{\bar x \tau_1 \tau_2}{x(\tau_1+\tau_2)} \right ] 
\end{eqnarray}
gives the Borel model for the magnetic form factor $G_M(Q^2)$.
Using  $G_M(Q^2) = F_1(Q^2)+F_2(Q^2)$  we  obtain 
\begin{eqnarray} 
\Phi_2(\tau_1,\tau_2,Q^2) 
= \frac{1}{(2\pi)^4(\tau_1+\tau_2)^3}  
\int_0^1 dx \, (2e_u-e_d)\ \bar x^3  
 \exp \left [ - Q^2\frac{\bar x \tau_1 \tau_2}{x(\tau_1+\tau_2)} \right ] 
\end{eqnarray}
for the double Borel transform of the amplitude related 
to the $F_2(Q^2)$ form factor. 
 The integrands of these representations 
 can be treated (for $\tau_1=\tau_2=\tau$) as models for the corresponding nonforward 
 parton densities. In the $Q^2=0$ limit, one obtains 
 models for  forward  parton densities.
 The use of local duality
 changes the exponential factor 
 $\exp [ - \bar x Q^2 \tau/ 2x ]$ into some  function 
 of $ \bar x Q^2 / xs_0$  without changing the pre-factors, i.e.,
    forward  parton densities.  
 Let us discuss main features of the models for forward densities.
 
 $\bullet $ In case of $\Phi_1$, the forward densities correspond to
 usual
 parton densities. The model gives:
 $$
  f_u^{\rm mod}(x) = 4 \bar x^2(3-2\bar x) = 4(1-x)^2(1+2x) \  \  \ ,
  \  \  \ 
  f_d^{\rm mod}(x) = 4 \bar x^3= 4(1-x)^3 \ .
  $$
 Just like in the pion case, the  
 condensate corrections and then DGLAP evolution
 will shift the distributions towards 
 smaller $x$. One may expect that these effects will
 modify both distributions in a similar way.
 So, it is interesting to compare relative shapes
 of the model $u$ and $d$ distributions.
 The main feature is that $d$ distribution
 has an extra power of $(1-x)$ compared to $u$ 
 distribution. The extra power of $(1-x)$ 
 in  $d$ distribution is 
 a well known phenomenological observation.
 For comparison,  GRV parametrization for 
  normalization point  $\mu^2 \sim 1$\,GeV$^2$
 can be rather accurately reproduced by \cite{Radyushkin:1998rt}
 $$f_u^{\rm phen}(x)=1.89 x^{-0.4} (1-x)^{3.5} (1+6x) \  \  \ ,
  \  \  \ 
 f_d(x)^{\rm phen}=0.54 x^{-0.6} (1-x)^{4.2} (1+8x) \ ,$$
 with $d$ distribution having extra power $(1-x)^{0.7}$ for 
 $x\to 1$.
It is easy to check that   the ratios \\
$f_d^{\rm mod}(x)/f_u^{\rm mod}(x)$ 
and  $f_d^{\rm phen}(x)/f_u^{\rm phen}(x)$ are very close to each other.

$\bullet $ In case of $\Phi_2$, the densities $e_a(x)$ 
correspond to the forward limit of the spin-flip GPDs
$$e_a(x)=E_a(x, \xi =0, t=0) \ .$$ 
They are inaccessible in deep inelastic scattering 
and other inclusive processes. 
However, the normalization integrals $\kappa_a$
for these  functions are related to the anomalous
magnetic moments of the nucleons: 
$$
e_u\kappa_u+e_d\kappa_d = \kappa_p \ \ \ , 
 \ \ \ e_d\kappa_u+e_u\kappa_d = \kappa_n \,  , 
 $$
and, hence, they are known: 
$$\kappa_u =2 \kappa_p + \kappa_n = 1.65\ \ \ , 
 \ \ \  \kappa_d =2 \kappa_n + \kappa_p= -2.07 \ .
 $$
 The model densities are
 $$
 e_u^{\rm mod}(x) =  8(1-x)^3\  \  \  ,\  \  \ 
 e_d^{\rm mod}(x) =  -4(1-x)^3  \  .
 $$
 The normalization integrals 
 for these functions are
 $\kappa_u^{\rm mod}=2$ and $ \kappa_d^{\rm mod} =-1$.
 The correct nontrivial sign of $\kappa_d$  is a very encouraging 
  indication that the model is a reliable  starting 
 point. 
 Another feature of the  model $e_{u,d}(x)$ distributions is that they both
 have an extra power of $(1-x)$ compared to $f_u(x)$.
 As a result, the  $F_2(Q^2)$ form factor, as we will see later,
 decreases faster with $Q^2$   than $F_1(Q^2)$ and $G_M(Q^2)$,
 again in agreement with experimental observations.

\section{Form factors at large momentum transfer}

In the Borel model, the  form factors are   given by the integrals 
 \begin{equation}
F^B(Q^2) =\int_0^1 dx \, f^B(x) \exp \left [ -Q^2 \tau \frac{1-x}{2x} \right ]
 \label{FFBorel}
 \end{equation}
involving the model parton density $f^B(x)$ and the exponential 
factor $e^{-(1- x) Q^2\tau/2x}$.
Hence,  at large $Q^2$, 
the form factors are  dominated by 
integration over  the region where 
$Q^2 \tau \bar x \sim 1$ or $1-x \sim  \tau/Q^2$.
For the local duality model, the whole 
integration region over $x$ is restricted 
to $(1+4s_0/Q^2)^{-1} \leq x \leq 1$, i.e.,
$1-x<4s_0/Q^2$ for large $Q^2$.  

Hence, the result of integration in both models is completely
determined by the behavior of parton densities at $x$ 
close to 1.
Namely, if 
$f(x)\sim (1-x)^{\nu}$ for $x\to 1$, then 
the relevant form factor drops like $1/(Q^2)^{\nu +1}$ at large $Q^2$.
This gives $1/Q^4$ for the asymptotics of 
the pion form factor,
$1/Q^6$ for the large-$Q^2$ behavior of  the nucleon form factors
$F_1(Q^2)$
and $G_M (Q^2)$,  and $1/Q^8$ for $F_2(Q^2)$.
All these  results seem to be in contradiction with 
the experimentally established exponents of the power-law 
behavior of  these form factors,
so one may be tempted to  conclude 
that our   models have no  chance to describe  the data. 

In fact, as  already mentioned, the local duality
prediction (\ref{LDpion}) is in excellent agreement with the results
of recent JLab data. In the nucleon case, the local duality
calculation of $G_M^p(Q^2)$ performed in 
 Ref. \cite{Nesterenko:1983ef,Nesterenko:1983gi}
 agrees with the data up to $Q^2 \sim 20$\,GeV$^2$
 (see  also Ref. \cite{Isgur:1984jm}, where the curve based on 
 Refs.\cite{Nesterenko:1983ef,Nesterenko:1983gi} 
 is compared with the results of other 
 approaches).
The ratio $F_2^p(Q^2)/F_1^p(Q^2)$ as calculated in the local duality model
agrees with the data based on Rosenbluth 
separation \cite{Andivahis:1994rq}, though at the highest $Q^2$ it 
is somewhat lower than the results of 
the polarization transfer experiments \cite{Gayou:2001qt}.
In all cases, deviations do not exceed 20-30\%. 
This means  that the model  curves mimic 
the  ``canonic'' $Q^2$  behavior of these 
form factors ($1/Q^2$ for $F_\pi(Q^2)$, $1/Q^4$ for $F_1^p(Q^2)$ and
 $G_M^p(Q^2)$, and  $1/Q^6$ for $F_2^p(Q^2)$).

The resolution of the paradox 
 is based on a trivial   observation that 
the model curves  are more complicated functions
than  just  pure 
powers of $1/Q^2$. 
 Their nominal  large-$Q^2$ asymptotics 
is achieved only at very large values of $Q^2$, 
well beyond the accessible region.  
 Thus,  conclusions made 
on   the basis of asymptotic relations
might  be of little importance in practice: a curve with  ``wrong'' 
 large-$Q^2$ behaviour might be quite successful phenomenologically
 in a rather wide range of $Q^2$.

\section{Improved Ans{\"a}tze}

The models discussed above, of course, have some drawbacks.
In particular,  the small-$x$ behavior 
of the model parton densities  is unsatisfactory: it does not have 
the 
 standard Regge-type behavior  $f(x)|_{x\to 0} 
 \sim x^{-\alpha(0)}$. 
 The excuse is that such a behavior cannot result from a single 
  lowest-order 
 triangle diagram: to get it, one should add an infinite number 
 of diagrams with the ladder structure in the $Q^2$ channel,
 and perform summation of all contributions.  
 A simple solution of this hopeless problem is 
 to take experimental forward distributions in Eq. (\ref{FFBorel}) instead 
 of the model ones. 
 Such an  approach was successfully used in Ref.\cite{Bakulev:2000eb}
 for the pion form factor and in Refs. 
 \cite{Radyushkin:1998rt,Diehl:1998kh} for the proton $F_1^p(Q^2)$ form factor.    
 The modified Borel (or Gaussian) model  for $F_1^p(Q^2)$  was 
 able to  successfully 
 describe the data up to   $Q^2 \sim 10$\,GeV$^2$.

 Next question is about  the exponential factor.
The Regge picture suggests 
$x^{-\alpha (t)}$ behavior at small $x$ or   
\begin{eqnarray}
{\cal F} (x,t) = f(x)  x^{-(\alpha (t)- \alpha (0))}  
\end{eqnarray}
model for the nonforward densities.
  Assuming a linear  Regge trajectory 
with the slope  $\alpha^{\prime}$, one gets  
\begin{eqnarray}
{\cal F}^{R} (x,t) = f(x)  x^{-\alpha^{\prime} t} 
=f(x)e^{-\alpha^{\prime} t \ln x} \  .  
\end{eqnarray} 
This Ansatz  was already discussed  in Refs. \cite{Goeke:2001tz}
and \cite{Vanderhaeghen:2002pg}. 
It also provides finite  mean squared radii 
\begin{eqnarray}
\langle r^2\rangle  &\,=\,& -6 \, \alpha_1^{\prime} \,
\int _{0}^{1}dx \; 
f(x) \ln x \; ,
\label{eq:rms} 
\end{eqnarray}
 which are in good agreement with experimental values \cite{Vanderhaeghen:2002pg}.
 For large $t$, experimental data support  Drell-Yan (DY) 
relation \cite{Drell:1969km} and Bloom-Gilman 
duality \cite{Bloom:1970xb}.  According to  DY, 
 if the parton density
behaves like $(1-x)^{\nu}$, then the relevant 
form factor should decrease as $1/t^{(\nu+1)/2}$ for large $t$.
The simplest idea is to attach  an extra $(1-x)$ factor 
in the exponential \cite{Burkardt:2002hr}, i.e. to take the model    
\begin{eqnarray}
{\cal F}_a^{R^{\rm mod}} (x,t) = f_a(x)  x^{-\alpha_1^{\prime} (1-x)t}  \  .
\end{eqnarray}
To calculate $F_2$, we need  an Ansatz for the  spin-flip nonforward
parton densities ${\cal E}_a(x,t)$. One can  assume the same 
model  
$
{\cal E}_a^{R^{\rm mod}} (x,t) =  e_a(x)\,  x^{-\alpha_2^{\prime}(1-x) t}
$
as for ${\cal F}_a (x,t)$, 
with possibly a slightly different
slope $\alpha_2^{\prime}$. 
To model the forward magnetic densities  $e_a (x)$, we can use the
lesson from the triangle diagram calculation that $e_u (x)$
has an extra power of $(1-x)$ compared to $f_u(x)$.
In fact, one can take the extra factors in the form $(1-x)^{\eta_a}$ 
with $\eta_a$'s  being  fitting parameters. 
Within this approach, it is possible to get a 
rather good description of all four nucleon form factors \cite{GPRV} 
(see Ref. \cite{Diehl:2004cx}
for a similar analysis).
  
 \section{Discussion}
 
 In this paper, we discussed basics of an approach 
 that uses QCD sum rule ideas to   build models  
for generalized parton distributions.
The underlying idea is to consider three-point functions
in which the hadrons are represented by local currents
with appropriate quantum numbers.
The necessary nonlocality bringing in parameters 
having the meaning of the hadron size can be  introduced
in several ways: 
by nonzero virtualities $p_i^2$ of the momenta associated 
with the currents, by  taking the double Borel transformation
from $p_i^2$'s to $\tau_i$'s, or through the local duality
prescription within the duality  square $s_0\times s_0$.
In terms of the basic function, the spectral density $\rho (s_1,s_2,t)$,
these possibilities correspond to integration 
with different weights: $a)$  $(s_1-p_1^2)^{-1} (s_2-p_2^2)^{-2}$
(producing the original amplitude $T(p_1^2,p_2^2,t)$), $b)$ 
$\exp [-s_1\tau_1-s_2\tau_2]$ (producing the double Borel transform
$\Phi (\tau_1,\tau_2,Q^2)$), and $c)$ $\theta(s_1\leq s_0) \theta(s_2\leq s_0)$
(local duality prescription).
We have considered only the lowest approximation for $\rho (s_1,s_2,t)$
(${\cal O} (\alpha_s)$ terms for pion form factor were 
discussed in refs.{\cite{Bakulev:2000uh,Braguta:2004ck}). 
One of the expected effects of higher-order corrections
is the emergence of the Regge-type behavior at small $x$.
Since there is no doubt about this outcome, such a behavior
can be introduced in the model expressions
by using experimental parton densities
instead of those generated by the lowest order term.

In  higher order diagrams, one  would also obtain 
contributions corresponding to the leading large-$Q^2$
asymptotics of perturbative QCD, like the one-gluon-exchange 
term for the pion form factor.
All such higher-order contributions
are  suppressed by  $\alpha_s/\pi\sim 1/10$ factor per each extra loop.
Such a suppression is manifest 
in the local duality result (\ref{LDpion}) for the 
pion form factor. Note, that the local duality
prediction is in perfect agreement with the data
despite the fact that the ${\cal O} (\alpha_s)$ term
containing the hard gluon exchange is insignificant 
compared to the lowest order term.
For the nucleon form factors, the leading pQCD two-gluon exchange 
term has {\it a priori} suppression by a factor of 100,
so it is unlikely to be relevant at any accessible momentum transfer.

Another pQCD prediction is about the $x\to 1$ behavior 
of parton densities. In the nucleon case, the leading $(1-x)^3$ term
corresponds to four-gluon-exchange diagrams and $\sim 10^{-4}$ suppression
compared to the ${\cal O}(\alpha_s^0)$ term. The latter has 
$(1-x)^2$ behavior for $u$ quarks and $(1-x)^3$ for $d$ quarks.
Nonperturbative effects and DGLAP
evolution are undoubtedly capable to shift these densities towards the experimentally
observed shapes. Furthermore, there is no need to make extra efforts
to bring in the relative $(1-x)$ suppression of the $d$ 
density: it is present in the starting approximation.

The lowest order term also implies  a  faster fall-off of the $F_2(Q^2)$ form factor 
compared to $F_1(Q^2)$. This effect  results from  the  extra $(1-x)$ 
power of $e_u^{\rm mod} (x)$ compared to $f_u^{\rm mod} (x)$.
 It should be emphasized that, in general,  the large-$Q^2$ behavior 
 of form factors in  the ${\cal O}(\alpha_s^0)$  approximation 
 is completely governed  by Feynman mechanism, 
 i.e., by $x\sim 1$ integration, so that the   large-$Q^2$ behavior 
 of form factors is always determined 
 by the $x\to 1$ behavior of the parton densities.
 The specific correlation pattern between the $\nu$ power in $(1-x)^\nu$ and
 $n$ in $1/(Q^2)^n$ depends on the structure of a factor like
 $\exp[-Q^2 \tau (1-x)/2x)]$ in the Borel model. 
  As we discussed, it should be modified to 
   $\exp[Q^2 \alpha^{\prime} (1-x)\ln x]$
  to impose the $n=(\nu+1)/2$ correlation dictated by the Drell-Yan relation.
    In contrast, in pQCD the large-$Q^2$ behavior 
 of form factors is governed by configurations in 
 which all the valence quarks carry finite momentum fractions $x_i$,
 i.e., the $1/Q^4$ behavior of $F_1(Q^2)$ {\it is not a consequence}
 of the $(1-x)^3$ behavior of the parton densities. The 
 relation $n=(\nu+1)/2$ in pQCD is just an accidental  correlation  
 between two parameters. In other words, there is a correlation 
 between $\nu$ and $n$ in pQCD because  both are determined 
 by the same hard gluon exchange mechanism, but there is no causal
 connection between these two numbers.
 As noted in the pioneering paper 
 \cite{Lepage:1980fj},  there is no Drell-Yan/Feynman  mechanism in pQCD. 
 
Summarizing, the gross features of generalized parton distributions
are dominated by nonperturbative dynamics, and, hence, 
we need nonperturbative approaches to build models for GPDs.
The models  motivated by QCD sum rule ideas have 
already made several successful predictions, and they also have 
the advantage of being closely related to perturbative calculations,
which allows to  satisfy   nontrivial constraints 
imposed on  GPDs. This makes the QCD sum rule based approach  an attractive  
possibility for building realistic models of GPDs.

\begin{acknowledgement}

I use this opportunity to thank my friend and 
colleague Prof. Dr. Klaus  Goeke for 
(in)numerous invitations to  visit the Institute of 
Theoretical Physics II in Bochum and conferences in Germany; his    
warm hospitality, inspiring discussions and  support.
This paper was strongly influenced by  my contacts with the members 
of his group M.V. Polyakov, C. Weiss, P.V. Pobylitsa, N.G. Stefanis, W. Schroers,
N. Kivel,  
and visitors to Bochum: A.P. Bakulev, S.V. Mikhailov, R. Ruskov (all from Dubna),  
D.I.  Diakonov, V. Petrov, M. Praszalowicz, 
and M.M. Musakhanov.  I  also thank I.V. Musatov, A. Belitsky, V.M. Braun,
A. Schaefer, A. Freund, M. Vanderhaeghen, M.Guidal,
 M. Diehl, and P. Kroll for stimulating discussions of models 
for generalized parton distributions.
This work  is supported by the US 
 Department of Energy  contract
DE-AC05-84ER40150 under which the Southeastern
Universities Research Association (SURA)
operates the Thomas Jefferson Accelerator Facility, 
and by the Alexander von Humboldt Foundation.

\end{acknowledgement}

\end{document}